\documentclass[twocolumn,showpacs,preprintnumbers,amsmath,amssymb]{revtex4}
\usepackage{mathrsfs}
\usepackage{amsmath}
\usepackage{graphicx,epsfig}

\begin{document}

\title {Characterization of phase transition in Heisenberg mixtures from density functional theory}

\author{ L. S. Li  \footnote
{Electronic mail~:liliangsheng@itp.ac.cn} and X. S. Chen \footnote
{Electronic mail~:chenxs@itp.ac.cn}} \affiliation{Institute of
Theoretical Physics, Chinese Academy of Sciences, P.O. Box 2735,
Beijing 100080, China}

\begin{abstract}
The phase transition of hard-sphere Heisenberg and Neutral Hard
spheres mixture fluids has been investigated with the density
functional theory in mean-field approximation (MF). The matrix of
second derivatives of the grand canonical potential $\Omega$ with
respect to the total density, concentration, and the magnetization
fluctuations has been investigated and diagonalized. The zero of the
smallest eigenvalue $\lambda_s$ signalizes the phase instability and
the related eigenvector $\textbf{x}_s$ characterizes this phase
transition. We find a Curie line where the order parameter is pure
magnetization and a mixed spinodal where the order parameter is a
mixture of total density, concentration, and magnetization. Although
in the fixed total number density or temperature sections the
obtained spinodal diagrams are quite similar topology, the
predominant phase instabilities are considerable different by
analyzing $\textbf{x}_s$ in density-concentration-magnetization
fluctuations space. Furthermore the spinodal diagrams in the
different fixed concentration are topologically different.
\end{abstract}

\pacs{05.70.Jk, 61.20.Gy, 64.60.Cn, 61.25.Em}

\maketitle

\section{INTRODUCTION}

The study of the properties of the Heisenberg liquid emerged as one
of the fascinating theoretical problems since it was stimulated by
MC simulations  of Heisenberg fluid \cite{NW1995}, which strongly
hint that the ferromagnetic transition of the Heisenberg fluid
challenge the traditional viewpoint to ferromagnetic transition
theory. Nijmeijer \emph{et al.} found the value of critical
exponents of the Heisenberg and Ising fluid\cite{NPR1998} differ
from the expected results for the lattice models. They suggested
that the the effective exponents is related to Fisher
renormalization with the fixed density. Recently Mryglod \emph{et
al.} simulated the larger Heisenberg fluid system and obtained the
critical exponents from standard finite size scaling
theory\cite{MOF2001}. They argued the effective exponents is
corrected by Fisher renormalization when a thermdynamics system
under a constraint\cite{F1968}. Although the spin liquids exhibit
interesting phase transition by simulations, the formal theory of
critical phenomena has not solved this problem completely. In fact
the spin fluids have a complex phase behavior as the coupling
between the additional spin degrees and spatial coordinates. Besides
ordinary gas-liquid phase, such a spin fluid model displays
paramagnetic gas-ferromagnetic Liquid phase transitions, critical
end point, and tri-critical point. Many models have been proposed
such as the discrete Ising
\cite{NPR1998,WN1996,FK1998,K2001,FMO2007}, continuous XY
\cite{OFMF2005,LL2007} and Heisenberg fluid
\cite{NW1995,MOF2001,LWABS1994,TTTWN1995,WNTT1997,LL1998,LLW1998,LWT1998,LLC2008}.
Tavares \emph{et al}. using both a mean field (MF) and a more
refined modified mean field (MMF) density functional theory have
found in some regime mixed first-order transition, namely a
condensation-ordering transition\cite{TTTWN1995} in the pure
Heisenberg fluid. Both theoretical works and MC results showed that
the first-order transition (i.e., an isotropic vapor phase and a
ferromagnetic liquid phase) are the mixed transition of the ordering
fluctuations and density fluctuations and Li \emph{et al}.
investigated it using the method proposed by Chen \emph{et al.}
\cite{CKF1991,CF1992} to characterize the mixed phase
transition\cite{LLC2008}.

The phase diagrams of  binary spin mixture calculated by the mean
field theory and Monte Carlo simulation show many fascinating phase
behaviors which come from competition among magnetic, condensation
and concentration fluctuations\cite{FF2005}. So far the
microscopically motivated studies try to comprehend the picture of
phase behavior of these fluids. However, as we shall demonstrate
here, knowledge of the complete phase diagram is essential but not
enough. In order to understand the existence of the mixed
transition, we have to classify the phase transition characters. So
our purpose in the present work is to characterize the phase
transition along ordering-condensation-demixing transition line.
Therefore, we investigate a mixture of Heisenberg hard spheres  and
Neutral hard spheres (HHNH) using the method\cite{CKF1991,CF1992} to
characterize phase transition. We address these questions using
density functional theory in the so-called simple mean-field theory.
As a result, the phase transitions of different character take place
in this system, as studies of the phase diagram and of the
associated critical phenomena have been shown.

\section{MODEL}

In this paper we consider a binary mixture fluid of Heisenberg
Hard-Sphere (HHS) and Neutral Hard-Sphere (NHS) with equal diameters
$\sigma$. The pair potential for two of such particles at positions
$\textbf{r}_1$ and $\textbf{r}_2$ is given by

\begin{eqnarray}
u_{ab}({\bf r}_1,\omega_1,{\bf
r}_2,\omega_2)=\Delta(ab)u^{ss}(r,\omega_1,\omega_2)+u^{hs}(r),
\end{eqnarray}
where $r=|{\bf r}_1 - {\bf r}_2|$ is the distance between particles
at positions ${\bf r}_1$ and ${\bf r}_2$ and $\omega_i =
(\theta_i,\phi_i)$ are the orientations of particles with $0 \leq
\theta_i \leq \pi$ and $0 \leq \phi_i \leq 2\pi$. The subscripts $a$
and $b$ denote the components in fluid considered \{a(b)= HHS,
NHS\}, and we define

\begin{equation}
\Delta(ab)= \left\{\begin{aligned}
1,~~~&a=b=HHS, \\
0,~~~&a=b=NHS,~a\neq b,
\end{aligned} \right.
\end{equation}
which means all the particles have a diameter yielding a repulsive
hard sphere potential $u^{hs}$ and only the HHS particles attract
each other via a Heisenberg potential $u^{ss}$. The repulsive hard
sphere interaction and the spin part are given by
\begin{equation}
u^{hs}(r)=\left\{ \begin{aligned}
+\infty,~~~ &r\leq \sigma, \\
0,~~~ &r>\sigma.
\end{aligned} \right.
\end{equation}
and
\begin{equation}
u^{ss}(r,\omega_1,\omega_2)=\left\{ \begin{aligned}
0,~~~&r\leq \sigma, \\
-J(r){\bf s}_1\cdot{\bf s}_2,~~~&\sigma < r \leq r_c\\
0,~~~&r>r_c.
\end{aligned} \right.
\end{equation}
where
\begin{eqnarray}
J(r)=\epsilon\frac{e^{-z(r/\sigma-1)}}{r/\sigma},
\end{eqnarray}
here ${\bf s}_i$ is a unit vector ($|\textbf{s}_i|=1$) in the
direction of the spin moment with

\begin{equation}
{\bf s}_1\cdot{\bf s}_2=\cos(\omega_1,\omega_2)=\cos\theta_1
\cos\theta_2 + \sin\theta_1 \sin\theta_2 \cos(\phi_1-\phi_2).
\end{equation}

In this paper we set the cutoff distance of the Yukawa potential
$r_c$ for $r_c=\infty$ and the dimensionless parameter $z$ for
$z=1$. To have ferromagnetic phase favoring parallel orientations
the coupling constant $\epsilon$ here is taken to be positive.

\section{DENSITY FUNCTIONAL AND MEAN-FIELD APPROXIMATION}

 The grand potential free energy $\Omega$ of a
nonuniform spin liquid mixtures is the minimum of the functional,

\begin{eqnarray}
&&\Omega[{\rho_{H}({\bf r},\omega),\rho_{N}({\bf r}),V,T}]=
F[\rho_{H}({\bf r},\omega),\rho_{N}({\bf r})]\\\nonumber &+&\int
d\omega d{\bf r} \rho_{H}({\bf r},\omega)(V_{ext}^H({\bf
r},\omega)-\mu_H)\\\nonumber &+&\int d{\bf r} \rho_{N}({\bf
r})(V_{ext}^N({\bf r})-\mu_N), \label{fall}
\end{eqnarray}
where index H and N means HHS and NHS, respectively. $\mu$ is the
chemical potential and the intrinsic Helmholtz free energy of the
inhomogeneous fluid $F[\rho_{H}({\bf r},\omega),\rho_{N}({\bf r})]$
is a unique functional of the densities $\rho_{H}({\bf r},\omega)$
and $\rho_{N}({\bf r})$. The Helmholtz free energy can be written as
\cite{R1979}

\begin{eqnarray}
&&F[\rho_{H}({\bf r},\omega),\rho_{N}({\bf r})]=F^{hs}[\rho_H({\bf
r},\omega),\rho_N({\bf
r})]\\\nonumber&+&\frac{1}{2}\int_0^1\:d\lambda \int \:d{{\bf
r}_1}d{{\bf r}_2}d\omega_1d\omega_2 g({\bf r_1},\omega_1,{{\bf
r}_2},\omega_2;\lambda)\\\nonumber &&\rho_{H}({\bf
r_1},\omega_1)u^{ss}(r,\omega_1,\omega_2) \rho_{H}({{\bf
r}_2},\omega_2)\label{helmholtz}
\end{eqnarray}
where $F^{hs}[\rho_H({\bf r},\omega),\rho_N({\bf r})]$ is the
Helmholtz free energy of Hard-Sphere (HS) mixtures system and
$g({\bf r_1},\omega_1,{{\bf r}_2},\omega_2;\lambda)$ is the pair
distribution function in a system which the particles interact via a
pairwise potential

\begin{equation}
u_{\lambda}(r,\omega_1,\omega_2)=h^{hs}(r)+\lambda
u^{ss}(r,\omega_1,\omega_2).
\end{equation}

The density of Heisenberg particle $\rho_{H}({\bf r},\omega)$ can be
split into the number density $\rho_H({\bf r})$ and a normalized
factor $O_H({\bf r},\omega)$ as

\begin{equation}
\rho_H({\bf r}, \omega)=\rho_H({\bf r})O_H({\bf r},\omega)
\end{equation}
where

\begin{equation}
\int d\omega O_H({\bf r},\omega) = 1
\end{equation}

In the local density approximation the Helmholtz free energy of HS
mixtures system\cite{CS1969}

\begin{eqnarray}
F^{hs}&=&\frac{1}{\beta}\int d{\bf r}d\omega \rho_H({\bf
r},\omega)\left[\ln(\Lambda^3_H4\pi \rho_H({\bf
r},\omega))-1\right]\\\nonumber&+&\frac{1}{\beta}\int d{\bf
r}\rho_N({\bf r})\left[\ln(\Lambda^3_N4\pi \rho_N({\bf
r}))-1\right]\\\nonumber&+&\frac{1}{\beta}\int
d\textbf{r}\rho(\textbf{r})
\frac{4\eta(\textbf{r})-3\eta(\textbf{r})^2}{(1-\eta(\textbf{r}))^2}
\end{eqnarray}
where $\beta =1/k_B T$ is the inverse temperature, $\Lambda_a$ is
the thermal de Broglie wavelength of species $a$, the total density
$\rho({\bf r})=\rho_H({\bf r})+\rho_N({\bf r})$ and $\eta ({\bf
r})=(\pi/6)\rho ({\bf r}) \sigma^3$ is the packing fraction.

In the mean field approximation where the pair distribution function
takes its large-distance limit $g({\bf r_1},\omega_1,{{\bf
r}_2},\omega_2;\lambda)=1$ \cite{TTTWN1995}, the part of the free
energy related to the spin-spin interactions in Eq.(\ref{helmholtz})
becomes

\begin{eqnarray}
F^{ss}_{MF}=\frac{1}{2}\int d{{\bf r}_1}d{{\bf
r}_2}d\omega_1d\omega_2 \rho_{H}({\bf r_1},\omega_1)
u^{ss}(r,\omega_1,\omega_2) \rho_{H}({{\bf r}_2},\omega_2)
\end{eqnarray}

After the decomposition $\rho_H ({\bf r},\omega)=\rho_H ({\bf
r})O_H({\bf r},\omega)$, the minimum condition of the functional
$\Omega[\rho_H({\bf r},\omega), \rho_N({\bf r})]$ is equivalent to
the simultaneous minimization of the grand canonical functional with
respect to the  number densities
\begin{eqnarray}
\frac{\delta \Omega [\rho_H({\bf r},\omega), \rho_N({\bf
r}),T,\mu]}{\delta \rho_H({\bf r})}=0,\\\nonumber\frac{\delta \Omega
[\rho_H({\bf r},\omega), \rho_N({\bf r}),T,\mu]}{\delta \rho_N({\bf
r})}=0,\label{equ1}
\end{eqnarray}
and the orientational configuration
\begin{equation}
\frac{\delta \Omega [\rho_H({\bf r},\omega), \rho_N({\bf
r}),T,\mu]}{\delta O_H({\bf r},\omega)}=0.\label{equ2}
\end{equation}

In the absence of the external field ($V_{ext}^H({\bf
r},\omega)=V_{ext}^N({\bf r})=0$) the system is homogeneous in
position, but it could be ordered in orientation. So we have $\rho_H
({\bf r})=\rho_H$, $\rho_N ({\bf r})=\rho_N$ and $O_H ({\bf
r},\omega)=O_H (\omega)$. In the mean field approximation we obtain
the bulk expression of the grand-canonical free energy density

\begin{eqnarray}\label{mfgf}
&&\frac{1}{V}\Omega_{MF}[O_H(\omega),\rho_H,\rho_N,T,\mu]
=f_{CS}^{hs}(\rho_H,\rho_N)\\\nonumber&-&\frac{1}{2}J^{int}\rho^2_H
|\bar{{\bf s}}|^2+\frac{\rho_H}{\beta} \int d \omega O_H (\omega)\ln
[4\pi O_H (\omega)]-\mu_H \rho_H-\mu_N\rho_N,
\end{eqnarray}
here the Helmholtz free energy density of hard-sphere system
$f_{CS}^{hs}(\rho_H,\rho_N)$ is given by Carnahan and Starling
\cite{CS1969}
\begin{eqnarray}
f_{CS}^{hs}(\rho_H,\rho_N)=\frac{\rho_H}{\beta}\left[\ln (\rho_H
\Lambda^3_H)-1\right]\\\nonumber +\frac{\rho_N}{\beta}\left[ \ln
(\rho_N
\Lambda^3_N)-1\right]+\frac{\rho}{\beta}\frac{4\eta-3\eta^2}{(1-\eta)^2}
\end{eqnarray}
and
\begin{eqnarray}
\bar{{\bf s}}&=& \int d \omega  O_H (\omega)\; {\bf s}(\omega),\\
J^{int}&=& \int_\sigma^{\infty}\:d r 4\pi r^2
J(r)=8\pi\varepsilon\sigma^3.
\end{eqnarray}

From the equilibrium condition (\ref{equ2}) and the mean field
grand-canonical free energy (\ref{mfgf}) we can obtain the
equilibrium orientational distribution function

\begin{equation}
O_H(\omega)=\frac{e^{\beta \rho_H J^{int} {\bf s} \cdot \bar{{\bf
s}}}}{\int d\omega e^{\beta \rho_H J^{int} {\bf s} \cdot \bar{{\bf
s}}}}.
\end{equation}

Here we restrict to the case $O_H(\omega) =O_H(\theta)$ and have
\begin{equation}
O_H(\omega)=O_H(\theta)=\frac{1}{4\pi} e^{k(x)+x\cos\theta},
\end{equation}
where
\begin{eqnarray}
&&k(x)= \ln(x/\sinh x),\\
&&x=\beta\rho_HJ^{int}\xi,
\end{eqnarray}
with the average magnetization $\xi =\int d \omega O_H (\theta)
\cos\theta $ determined by
\begin{equation}
\xi = \coth (\beta\rho_H J^{int}\xi)-\frac{1}{\beta\rho_H
J^{int}\xi}.\label{magnetization}
\end{equation}

Then the grand function free energy density in an equilibrium state
is expressed by
\begin{eqnarray}
&&\frac{1}{V}\Omega_{MF}[O_H(\theta),\rho_H,\rho_N,T,\mu]
=f_{CS}^{hs}(\rho_H,\rho_N)\\\nonumber&+&\frac{1}{2}J^{int}\rho^2_H
\xi^2+\frac{\rho_H}{\beta} k(x)-\mu_H \rho_H-\mu_N \rho_N.
\end{eqnarray}

\section {The SPINODAL OF HEISENBERG HARD MIXTURE LIQUIDS }

\begin{figure}[!h]
\centering
\includegraphics[width=1.0\columnwidth]{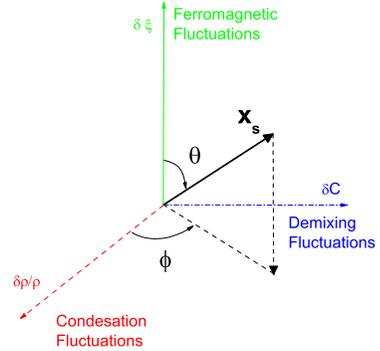}
\caption{(Color Online) A schematic plot of the eigenvector which
characterizes the phase transition. In the space $( \delta
\rho/\rho,\delta C,\delta \xi)$ the eigenvector can be described by
angles $\phi$ and $\theta$ which is in the range
$\theta\in(-90^0,90^0]$ and $\phi\in(-180^0,180^0]$. When $\theta=0$
the phase transition is a pure ferromagnetic phase transition. The
phase transition is pure condensation when $\theta = 90^0$ and $\phi
= 0$. The phase transition is pure demixing when $\theta = 90^0$ and
$\phi = 90^0$.}\label{co}
\end{figure}

At a stable equilibrium state the grand potential has its minimum
and its variation with respect to the changes of number densities
and magnetization should be positive \cite{CKF1991},
\begin{eqnarray}
\delta \Omega/V &=& \frac{1}{2V}\Bigg[\frac{\partial^2
\Omega}{\partial \rho^2_H}(\delta\rho_H)^2+\frac{\partial^2
\Omega}{\partial\xi^2}(\delta\xi)^2\\\nonumber&+&\frac{\partial^2
\Omega}{\partial \rho^2_N}(\delta\rho_N)^2+2\frac{\partial^2
\Omega}{\partial \rho_H\partial \xi}\delta\rho_H
\delta\xi\\\nonumber&+&2\frac{\partial^2 \Omega}{\partial
\rho_N\partial \xi}\delta\rho_N \delta\xi+2\frac{\partial^2
\Omega}{\partial \rho_H\rho_N \xi}\delta\rho_H \delta\rho_N\Bigg]>0.
\end{eqnarray}

The variation of the grand potential can be rewritten in a matrix
form
\begin{equation}\label{matrix}
\delta \Omega /V = \frac{1}{2}
 \left(
  \begin{array}
 {ccc}
\delta\rho_H & \delta\rho_N &\delta \xi
  \end{array}
 \right)
 \left(
\begin{array}
 {ccc}
M_{HH} & M_{HN} & M_{H\xi}\\
M_{NH} & M_{NN} & M_{N\xi}\\
M_{\xi H} & M_{\xi N} & M_{\xi\xi}
\end{array}
\right) \left(
  \begin{array}
 {c}
\delta\rho_H \\
\delta\rho_N \\
\delta \xi
  \end{array}
 \right).
\end{equation}

We want to understand the character of the phase transition, which
combinations of condensation, phase separation, and ferromagnetic
order fluctuations are leading to the phase transition. For the
following analysis it is more convenient to rewrite the matrix in
terms of the total density $\rho=\rho_H+\rho_N$ , the concentration
$C=\rho_H/\rho$, and the ferromagnetic ordering, $\xi$. We reexpress
the variation of the grand potential in new fluctuations space
($\delta\rho,\delta C,\delta \xi$)

\begin{equation}\label{matrix}
\delta \Omega /V = \frac{1}{2}
 \left(
  \begin{array}
 {ccc}
\delta\rho/\rho& \delta C &\delta \xi
  \end{array}
 \right)
 \left(
\begin{array}
 {ccc}
M_{\rho\rho} & M_{\rho C} & M_{\rho\xi}\\
M_{C\rho} & M_{CC} & M_{C\xi}\\
M_{\xi\rho} & M_{\xi C} & M_{\xi\xi}
\end{array}
\right) \left(
  \begin{array}
 {c}
\delta\rho/\rho \\
\delta C \\
\delta \xi
  \end{array}
 \right).
\end{equation}
where total one particle density fluctuations
$\delta\rho=\delta\rho_H+\delta\rho_N$, the demixing fluctuations
$\delta C=\delta \frac{\rho_H}{\rho}
=\rho^{-2}[\rho_N\delta\rho_H-\rho_H\delta\rho_N]$. And the matrix
$\textbf{M}$ is defined by

\begin{equation}\label{matrix}
\textbf{M}=
 \left(
\begin{array}
 {ccc}
M_{\rho\rho} & M_{\rho C} & M_{\rho\xi}\\
M_{C\rho} & M_{CC} & M_{C\xi}\\
M_{\xi\rho} & M_{\xi C} & M_{\xi\xi}
\end{array}
\right)
\end{equation}
which has three eigenvalues $\lambda_1,\lambda_2,\lambda_3$ with the
corresponding eigenvectors $\textbf{x}_1,\textbf{x}_2,\textbf{x}_3$.
The positive eigenvalues $\lambda_1$, $\lambda_2$, and $\lambda_3$
of a stable state guarantee that the grand free energy will increase
with respect to any variation of total number density,
magnetization, and concentration. If the smallest eigenvalue
$\lambda_s$ vanishes, the system can deviate away from the original
state without any increase of the grand free energy and becomes
unstable, which is a phase transition. The eigenvector
$\textbf{x}_s$ corresponding to the zero eigenvalue characterizes
this phase transition precisely \cite{CKF1991,CF1992} and
$\textbf{x}_s$ is the order parameter. Therefore, We investigate the
direction of eigenvector {${\bf x}_s$} in the ($\delta\rho,\delta
C,\delta \xi$) fluctuations space, which can characterize the type
of phase transitions.

In Fig.\ref{co} we show a schematic plot of the Euclidean
eigenvector ${\bf x}_s$ in the space $(\delta\rho/\rho,\delta
C,\delta \xi)$ with unit vectors $({\bf e}_x,{\bf e}_y,{\bf e}_z)$
 of the Cartesian  system. We normalize the eigenvector $|{\bf
x}_s|=1$ of the zero eigenvalue, and calculate the angles $\theta$
and $\phi$ of ${\bf x}_s$ in the spherical coordinate
\begin{eqnarray}
&&\theta=\arccos({\bf x}_s\cdot{\bf e}_z),\\
&&\phi=\arccos\left(\frac{{\bf x}_s\cdot{\bf e}_x}{\sqrt{(1-({\bf
x}_s\cdot{\bf e}_z)^2)}}\right).
\end{eqnarray}

The angle $\phi$ and $\theta$ describe the portion of the total
number density, the concentration and the magnetization in the phase
transition and is defined in the range $-90^0< \theta \le 90^0,
-180^0<\phi\le180^0$. In general, the phase transition of the
Heisenberg fluid is a combination of condensation and ferromagnetic
phase transition. If $\theta=0$, we have a pure ferromagnetic phase
transition. For $\theta$ near zero and $\phi\neq0$, we have a
ferromagnetic dominant phase transition accompanied by a weak
condensation and demixing fluctuations. If $\theta=90^0$ and
$\phi=0$, we have then a pure gas-liquid phase transition. For
$\theta=90^0$ and $\phi$ near zero, we have a condensation dominant
phase transition accompanied by a weak phase separation and
ferromagnetic phase transition. If $\theta=90^0$ and $\phi=90^0$, we
have then a pure demixing phase transition. For $\theta=90^0$ and
$\phi$ near $90^0$, we have a demixing dominant phase transition
accompanied by a weak ordering and condensation phase transition.
Now we investigate the phase behavior of the binary fluid by
determining the border of stable region from the zero point of the
smallest eigenvalue $\textbf{x}_s$. What we will obtain is the
so-called spinodal.

\section{RESULTS AND DISCUSSION}

\begin{figure}[!h]
\centering
\includegraphics[width=1.0\columnwidth]{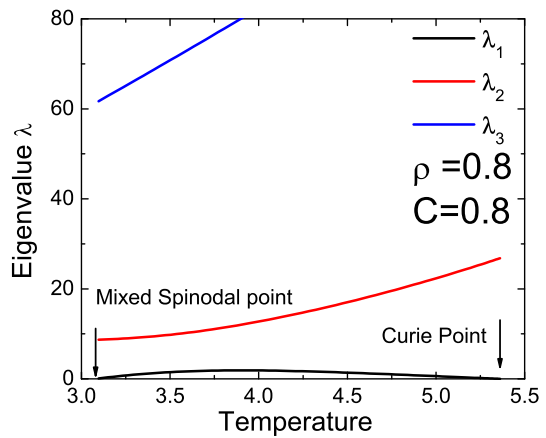}
\caption{(Color Online) Plot of the eigenvalues $\lambda_1$,
$\lambda_2$,and $\lambda_3$, vs $T$ for the Heisenberg Hard sphere
mixture fluid at the density $\rho^*=0.8$ and the concentration
$C=0.8$. The Curie point and the mixed spinodal point here are on
the Curier line and the mixed spinodal of Fig.
\ref{RhoCons}(f).}\label{L}
\end{figure}

\begin{figure}
\centering
\includegraphics[width=1.3\columnwidth]{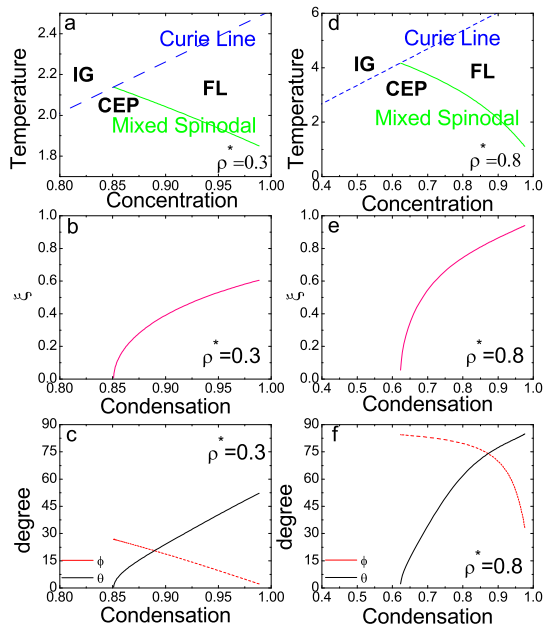}
\caption{(Color Online) (a,d) The spinodal curve of the Heisenberg
Hard Sphere mixture fluid in different density sections
($\rho^*=0.3, 0.8$). The blue dashed line, the Curies Line; the
green solid line, the mixed spinodal curve; CEP, critical end point;
IG, Isotropic Gas; FL, Ferromagnetic Liquid. (b,e) The magnetization
$\xi$ along the mixed spinodal curve with respect to concentrations.
The pink solid line, $\xi$. (c,f) The angles along the mixed
spinodal curve with respect to concentrations. The black solid line,
the angle $\theta$ ; the red dashed line, the angle
$\phi$}\label{RhoCons}
\end{figure}

In Fig.\ref{L} the eigenvalues $\lambda_1$, $\lambda_2$ and
$\lambda_3$ are shown as a function of the reduced temperature
$T^*=1/\beta\epsilon$ for the reduced density
$\rho^*=\rho\sigma^3=0.8$ and the concentration $C=0.8$. The
eigenvalues $\lambda_2$ and $\lambda_3$ always keep to be positive,
but the the smallest eigenvalue $\lambda_s=\lambda_1$ approaches
zero when decreasing or increasing the temperature. The instability
at higher temperature is actually on the Curie line where
$\theta=0$. And the smallest eigenvalue at lower temperature meet
the mixed spinodal line where $\theta=62^0$ and $\phi=80^0$. So we
have a dominant phase separation accompanied by a weaker ordering
and weakest gas-liquid phase transition. With the zero points of
$\lambda_s$ for different concentrations $C$ we can get the spinodal
phase diagram of the Heisenberg Hard sphere liquid, which is shown
in Fig.\ref{RhoCons}.

\subsection{Case: In the isotropic phases $\xi=0$}

At the isotropic phase of the Heisenberg fluid, there is no total
magnetization and $x=0$. In this case we have
\begin{equation}
\lim_{x\to 0}\frac{\partial^2 k(x)}{\partial x^2}=-\frac{1}{3}
\end{equation}

For all densities $\rho$ the element $M_{ij}$ is positive expect the
element $M_{\xi\xi}$, which can become zero and $\theta = 0$
corresponds to a pure ferromagnetic phase transition.

From $M_{\xi\xi}=0$ we can get the Curie line of the Heisenberg Hard
mixture fluids. With the reduced density $\rho^*=\rho\sigma^3$ and
the reduced temperature $T^*=1/\beta\varepsilon$ we can express the
Curie line as
\begin{equation}
T^*=\frac{8\pi}{3}\rho^*
\end{equation}
which agrees with the result of Li \emph{et al} \cite{LLC2008}
because of the simple mean field approximation. And the mean-field
approximation that makes the phase transition be pure ferromagnetic.
In a more accurate theory the phase transition is not pure
ferromagnetic and should be accompanied by the weak condensation and
demixing phase transition.

\subsection{Case: $\xi\neq0$, the total density is fixed}

When $\xi \neq 0$ the determinant of the coefficient matrix $M$

\begin{equation}
 Det\
 \left[
\begin{array}
 {ccc}
M_{\rho\rho} & M_{\rho C} & M_{\rho\xi}\\
M_{C\rho} & M_{CC} & M_{C\xi}\\
M_{\xi\rho} & M_{\xi C} & M_{\xi\xi}
\end{array}
\right]=0.
\end{equation}
which can be solved numerically. In Fig.\ref{RhoCons} $(a)$, we show
the Curie line (Blue Dash Line) and the mixed spinodal (Green Solid
Line) of the HHNH mixtures with the fixed density $\rho^*=0.3$
section.

the magnetization and the angle as a function of concentrations. The
phase diagram Will be discussed in detail and characterized the type
of phase transitions in the method\cite{CF1992}.

In Fig.\ref{RhoCons} $(a)$ we obtain the spinodal phase diagram of
the Heisenberg Hard spheres mixture liquid in the constant density
in the temperature concentration plane. The mixed spinodal meets the
Curie line at the critical endpoint with the concentration $C=0.85$
and the reduced temperature $T^*=2.13$. Below the temperature of the
critical endpoint there is a first-order phase transition between
isotropic vapor and ferromagnetic liquid. The magnetization $\xi$
along the mixed spinodal is shown in Fig.\ref{RhoCons} $(b)$ by
using Eq.(\ref{magnetization}). The more Heisenberg particles the
the liquid have, it show the bigger the magnetization. For the mixed
spinodal the phase instability is a combination of condensation,
demixing and ferromagnetic phase transition. To characterize the
phase instability precisely we investigate the angles $\phi$ and
$\theta$ of the eigenvector ${\bf x}_s$ along it shown in
Fig.\ref{RhoCons} $(c)$. For concentration near $1.0$ the angle
$\phi$ is very small and positive and $\theta>45^0$, where the phase
instability is predominantly related to the condensation. The
positivity of $\phi$ means that an increase of the total number
density will be accompanied by a small demixing and ordering. This
result is plausible, because the increase of concentration (more
Heisenberg particles) will enhances the average attraction of the
system which results in more ordering and demixing. The
magnetization $\xi$ for the concentration larger than $0.95$ is
larger than $0.5$. With the decrease of the concentration the angle
$\phi$ increases but $\theta$ decreases continuously. When the
system approaches the critical endpoint $C=0.85$, the angle $\theta$
approaches $0$. The phase transition at critical endpoint is then a
pure ferromagnetic phase transition in the mean field approximation.

The spinodal curve and the CEP with $C=0.69$ and $ T^*=2.89$ at
$\rho^*=0.8$ section is displayed in Fig.\ref{RhoCons} $(d)$.
Corresponding the magnetization in Fig.\ref{RhoCons} $(e)$ and the
angles in Fig.\ref{RhoCons} $(f)$ have a obvious region
$0.735<C<0.965$ that $\theta>45^0$ and $\phi>45^0$, in which the
phase instability is predominantly related to the demixing phase
transition accompanied by the ordering and condensation. By
comparing Fig.\ref{RhoCons} $(c)$ with $(f)$ the demixing
predominant region disappear in the lower total density section.

\subsection{Case: $\xi\neq0$ and Temperature is fixed}

\begin{figure}[!h]
\centering
\includegraphics[width=1\columnwidth]{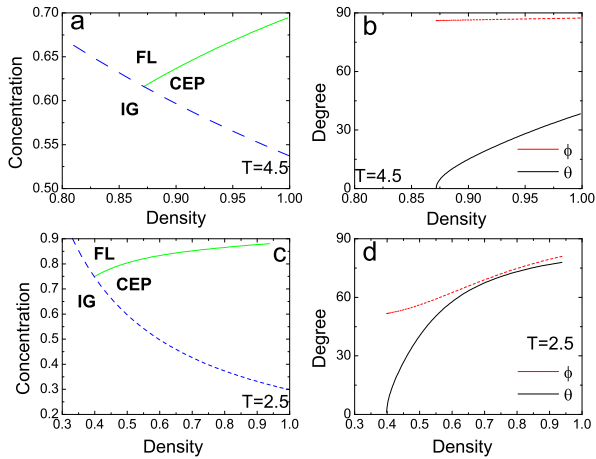}
\caption{(Color Online)(a,c) The spinodal curve of the Heisenberg
Hard Sphere mixture fluid in different Temperature sections ($T=4.5,
2.5$). The blue dashed line, the Curies Line; the green solid line,
the mixed spinodal curve; CEP, critical end point; IG, Isotropic
Gas; FL, Ferromagnetic Liquid. (b,d) The angles along the mixed
spinodal curve with respect to the total density. The black solid
line, the angle $\theta$ ; the red dashed line, the angle
$\phi$.}\label{TempCons}
\end{figure}

The spinodal diagrams in the density concentration plane are shown
in Fig. \ref{TempCons}(a,c). In order to elucidate the differences
of the mixture phase behavior at different temperatures, phase
behavior is characterized by $\theta$ and $\phi$. Fig.
\ref{TempCons}(b) show in the $T=4.5$ section the angles $\phi>85^0$
and $\theta<45^0$ along the mixed spinodal curve, in which the phase
instability is predominantly related to the ordering phase
transition accompanied by the weak demixing and tiny condensation.
But in the lower temperature section $T=2.5$ ( Fig.\ref{TempCons}(d)
) the instability predominantly related to the demixing accompanied
by the weak ordering and gas-liquid phase transition at $\rho*>0.6$.
Although the different temperature sections are quite similar
topology, the predominant phase instability is considerable
discrepancy between them. Recently, Fenz\emph{el at.} presented
concentration pressure phase diagrams of the ideal Ising mixture
fluids at different temperatures via the Gibbs ensemble Monte Carlo
simulation \cite{FF2005}, which the phase behavior can qualitatively
agree with the $T$ sections calculated in our model. On the other
hand the demixing phase transition in the mixtures can be easily
observed in the lower temperature section but ferromagnetic phase
transition predominate in the higher temperature section.

\subsection{Case: $\xi\neq0$ and Concentration is fixed}
\begin{figure}[!h]
\centering
\includegraphics[width=1.05\columnwidth]{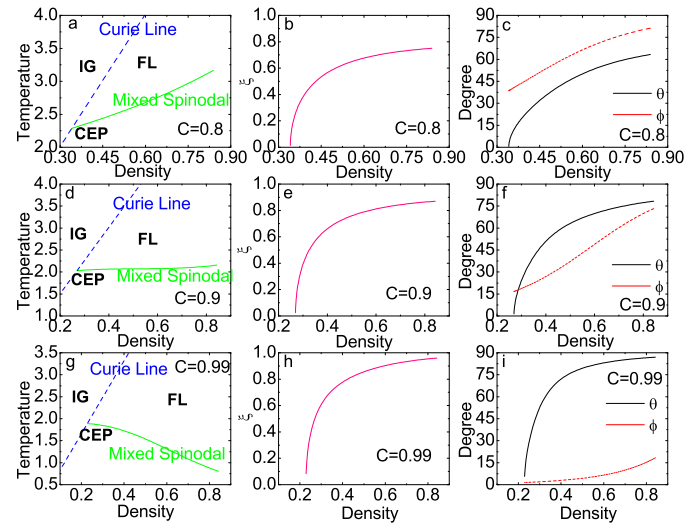}
\caption{(Color Online) (a,d,g) The spinodal curve of the Heisenberg
Hard Sphere mixture fluid in different Concentration sections
($C=0.8, 0.9, 0.99$). The blue dashed line, the Curies Line; the
green solid line, the mixed spinodal curve; CEP, critical end point;
IG, Isotropic Gas; FL, Ferromagnetic Liquid. (b,e,h) The
magnetization $\xi$ along the mixed spinodal curve with respect to
the total density. The pink solid line, $\xi$. (c,f,i) The angles
along the mixed spinodal curve with respect to the total density.
The black solid line, the angle $\theta$ ; the red dashed line, the
angle $\phi$.}\label{ConCons}
\end{figure}

Density-temperature spinodal diagrams are plotted in
Fig.(\ref{ConCons})(a,d,g) where the different sections belong to
different values of the parameter $C$. In Fig.\ref{ConCons} $(a)$
the critical endpoint is at $\rho^*=0.345$ and $T^*=2.28$ and the
slope of the mixed spinodal is positive which was found in
asymmetric binary dipolar mixtures\cite{RK2004}. In
Fig.\ref{ConCons} $(c)$ the phase instability is predominantly
related to the demixing in the region $\rho^*>0.55$ and the ordering
in the range of $0.345<\rho^*<0.55$. But the condensation
fluctuations is always weakest in the $C=0.8$ section. The positive
slope of the mixed spinodal means that increase of the total number
density, at fixed concentration, the first order phase transition
temperature increase with increasing $\rho^*$. When considering
mixtures at $C=0.9$ one finds from Fig. (\ref{ConCons})(d) that the
slope of the mixed spinodal move towards flat. Since increase of the
Heisenberg particle concentration implies that the condensation
instability gradually predominate the phase transition, as shown
Fig. (\ref{ConCons})(f). In the limit $C\rightarrow1.0$ one recovers
the MF spinodal diagram of the pure Heisenberg fluid contain only
two fluid phases, which is an isotropic vapor and a ferromagnetic
liquid\cite{LLC2008}. Here we set $C=0.99$, and then the spinodal
diagram is shown in Fig.\ref{ConCons} $(g)$ and the angle $\phi$ is
always near the zero and $\theta$ is large than $45^0$ when
$\rho*>0.25$ in Fig.\ref{ConCons} $(i)$. The predominant instability
is related to the condensation with small ordering and demixing
fluctuations which coincide with results of pure Heisenberg fluid.
In particular, the angle is obtained as a function of the
concentration in Fig.\ref{RhoCons} thereby providing an estimate of
the crossover between the demixing predominant and the condensation
predominant phase transition as expected. Such the crossover
phenomena can be precisely characterized by the method\cite{CF1992}.

\section{CONCLUSIONS}

In this paper we have investigated the phase transition of HHNH
mixtures by using the density functional theory in the mean-field
approximation. The phase instability of the system is discussed with
the method developed in Ref.(\cite{CF1992}). From the matrix of the
second derivatives of the grand canonical free energy $\Omega$ with
respect to the total particle density, the concentration and the
magnetization, we can determine the thermodynamic stability of the
system. When the smallest eigenvalue of the matrix becomes zero, the
system becomes unstable. The eigenvector corresponding the zero
eigenvalue can characterize the phase instability precisely. In the
total density, concentration, magnetization space $(\delta
\rho/\rho, \delta C, \delta \xi)$ the normalized eigenvector can be
described by two angles $\theta$ and $\phi$ which is in the range
$-90^0< \theta \le 90^0$ and $-180^0<\phi\le180^0$, respectively.

For temperature above of the critical endpoints, the angle $\theta =
0$ and the phase transition is pure ferromagnetic. This is the
result in the mean-field approximation. With a more accurate
approximation the angle $\theta$ does not equal exactly to $0$ and
is just near $0$. Then the total particle density and concentration
will be related to the phase transition, which is predominantly
ferromagnetic and accompanied by a weak condensation and demixing.
Below CEP there is a first-order phase transition between isotropic
gas and ferromagnetic liquid. The phase instability along the
spinodal near the ferromagnetic liquid is a combination of
condensation, demixing and ferromagnetic phase transition. The phase
behavior can topologically be worthy to be compared with the
simulations\cite{FF2005} in the different $T$ sections. Although the
different temperature sections are similar topology, in the lower
temperature section $T=2.5$ the instability predominantly is related
to the demixing but in the $T=4.5$ section the ordering phase
transition is predominant. The slope of the mixed spinodal is the
positive in the less concentration $C=0.8$ sections and is negative
in the $C=0.99$ section. Hence by accurately investigating the
angles we quantitatively describe that the phase instability
continuously changes from predominant demixing phase transition to
predominant gas-liquid phase transition. Although these crossover
phenomena in the concentration sections may be considered as
different phase transitions, our analysis shows that their origin in
fact is same.

These crossover phenomena within mean field approximation should be
corroborated Monte Carlo simulations and further explore the
combinations of phase transitions on the Curie line with more
accurate approximation.

\begin{acknowledgments}
This work was supported by the National Natural Science Foundation
of China under grant 10325418.
\end{acknowledgments}

\section{APPENDIX: THE ELEMENTS OF THE MATRIX $M$}
In this appendix we give the expressions of the matrix elements
(\ref{matrix}) in detail,
\begin{eqnarray}
M_{\rho\rho}&=&\rho^2\times\Big(C^2M_{HH}+2CM_{HN}
\\\nonumber&-&2C^2M_{HN}+(1-C)^2M_{NN}\Big),
\end{eqnarray}
\begin{eqnarray}
M_{CC}&=&\rho^2 (M_{HH}-2M_{HN}+M_{NN}),
\end{eqnarray}
\begin{eqnarray}
M_{\rho\xi}&=&M_{\xi\rho}=\rho C M_{H\xi},
\end{eqnarray}
\begin{eqnarray}
M_{C \xi}&=&M_{\xi C}=\rho M_{H\xi},
\end{eqnarray}
\begin{eqnarray}
 M_{C\rho}&=&M_{\rho
C}=\rho^2\times\Big(CM_{HH}\\\nonumber&+&(1-2C)M_{HN}-(1-C)M_{NN}\Big),
\end{eqnarray}
where the elements of the matrix are related to
\begin{eqnarray}
M_{HH}&=&\beta J_{int}^2\rho_H\xi^2 \frac{\partial^2k(x)}{\partial
x^2} + 2J_{int}\xi\frac{\partial k(x)}{\partial x}
+J_{int}\xi^2\\\nonumber &+&\frac{1}{\beta\rho_H}+
\frac{\partial^2\frac{\rho}{\beta}\frac{4\eta-3\eta^2}
{(1-\eta)^2}}{\partial\rho_H^2},\\
M_{NN}&=&\frac{1}{\beta\rho_N}+\frac{\partial^2\frac{\rho}
{\beta}\frac{4\eta-3\eta^2}{(1-\eta)^2}}{\partial\rho_N\partial\rho_N},\\
M_{\xi\xi}&=&\beta J_{int}^2\rho_H^3\frac{\partial^2k(x)}{\partial
x^2}+J_{int}\rho_H^2,\\
M_{HN}&=&M_{NH}=\frac{\partial^2\frac{\rho}{\beta}
\frac{4\eta-3\eta^2}{(1-\eta)^2}}{\partial\rho_H\partial\rho_N},\\
M_{H\xi}&=&M_{\xi H}=\beta J_{int}^2
\rho_H^2\xi\frac{\partial^2k(x)}{\partial x\partial
x}\\\nonumber&+&2J\rho_H\frac{\partial k(x)}{\partial x}+2J_{int}\rho_H\xi,\\
M_{N\xi}&=&M_{\xi N}=0.
\end{eqnarray}


\end{document}